\documentclass[onecolumn,epsfig,pre]{revtex4}
\usepackage{graphicx}
\usepackage{amssymb}
\usepackage{amsmath}
\usepackage{hyperref}
\usepackage{latexsym}

\def\be{\begin{equation}}
\def\ee{\end{equation}}
\def\bea{\begin{eqnarray}}
\def\eea{\end{eqnarray}}

\begin{document}

\title{Collective transport of weakly interacting molecular motors with Langmuir kinetics}
\author{Sameep Chandel$^{1}$, Abhishek Chaudhuri$^{1}$ and Sudipto Muhuri$^{2}$}

\affiliation{$^{1}$Indian Institute of Science Education and Research Mohali, Knowledge City, Sector 81, SAS Nagar - 140306, Punjab, India.\\
$^{2}$Department of Physics, Savitribai Phule Pune University, Ganeshkhind, Pune 411007, India}

\begin{abstract}
Filament based intracellular transport involves the collective action of molecular motor proteins. Experimental evidences suggest that microtubule (MT) filament bound motor proteins such as {\it kinesins} weakly interact among themselves during transport and with the surrounding cellular environment. Motivated by these observations we study a driven lattice gas model for collective unidirectional transport of molecular motors on open filament, which incorporates the short-range interactions between the motors on filaments and couples the transport process on filament with surrounding cellular environment through adsorption-desorption Langmuir (LK) kinetics of the motors. We analyse this model within the framework of a Mean Field (MF) theory in the limit of {\it weak} interactions between the motors. We point to the mapping of this model with the non-conserved version of  Katz-Lebowitz-Spohn (KLS) model. The system exhibits rich phase behavior with variety of inhomogeneous phases including localized shocks in the bulk of the filament. We obtain the steady state density and current profiles and analyse their variation as function of the strength of interaction. We compare these MF results with Monte Carlo simulations and find that the MF analysis shows reasonably good agreement as long as the motors are weakly interacting. We also construct the non-equilibrium MF phase diagram.
\end{abstract}

\maketitle

Motor proteins play a crucial role in various intracellular processes. Their collective action is involved in host of functions ranging from intracellular trafficking to the organization of the complex cytoskeletal network \cite{cell}.  Individual motor proteins such as {\it kinesin} and {\it myosin} hydrolyze ATP to actively move along the cellular filaments \cite{cell,howard}. While single molecule experiments have shed light on the mechano-chemical properties and functioning of individual motor proteins \cite{howard}, it has been established that many of the cellular functions such as cell division and intracellular transport is achieved by the collective effect of many motors. Many motors working in unison not only generates large forces that are required for many cellular processes but it is also responsible for long distance regulated transport inside the cell \cite{welte}. Thus it is of considerable interest to understand how the interaction between individual motor proteins affects the collective transport within the cell.

Experimental studies on transport of {kinesin} motors on MT filament suggests that the individual motors interact {weakly} with their neigbours on the MT and that this attractive interaction manifests as a propensity of the motors to cluster during their transport on cellular filaments \cite{roos,vilfan,seitz}. 
It has also been widely observed that spatial organization of the motors and the cellular cargo that they transport on the filaments is sensitive to the surrounding cellular environment. For example, experiments done to study the spatial distribution of pigment granules in extracts of {\it melanophore} cell has illustrated that altering the cellular environment by specific biochemical means affects the binding affinity of the motors to the filaments and which in turn can lead to spatial reorganization of the pigment granules \cite{borisy, gross, welte}. Further since cellular cargoes carried by the motors are loaded and offloaded at the filament ends, the role of the boundary processes at the filament ends also play a significant role in determining the spatial organization of the motors on individual filaments\cite{welte}.

\begin{figure*}[t] 
\begin{center}
\includegraphics[width=12cm]{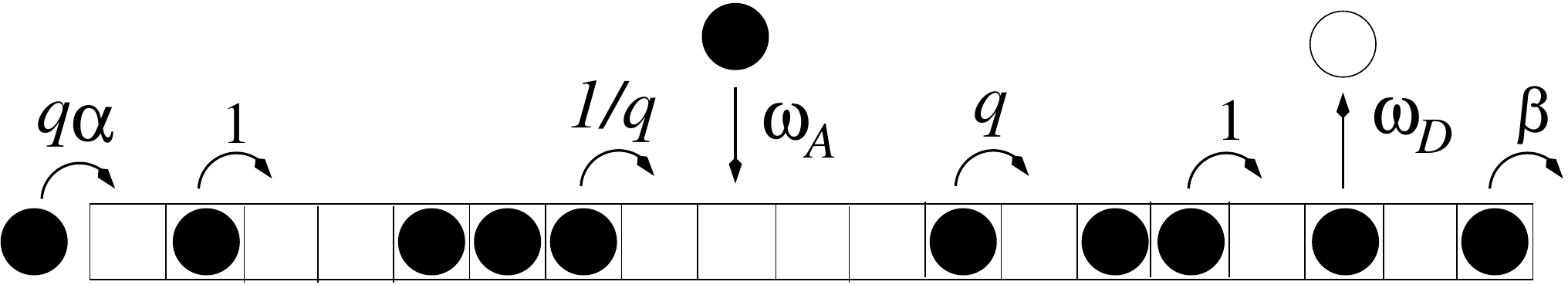} 
\caption{(color online)
Schematic diagram of the process.}
\label{3phase}
\end{center}
\end{figure*}
In light of these experimental finding, we study a minimal lattice gas model for unidirectional transport on a single filament which takes into account the weak interaction between the motors and their interaction with the surrounding environment at the microscopic scales within a simplified framework. The proposed framework provides a natural means to qualitatively understand some of the aspects of collective behaviour of motors and their spatial organization, that emerges at macroscopic cellular scales. We study this model within a Mean Field (MF) theory and compare the results with Monte Carlo simulation results.

The minimal model that we analyse, is a variant of the totally asymmetric exclusion process (TASEP) and belongs to the broad class of one-dimensional drive diffusive system \cite{schutz}. Driven diffusive systems in one dimension have been studied extensively from a theoretical perspective \cite{evans1,evans2,schutz,evans,popkov} and they exhibit a variety of interesting features and rich phase behaviour which includes boundary induced phase transitions \cite{evanssugden,freylet,freypre,evans,popkov}. Such boundary induced phase transitions in these driven systems are in contrast to one dimensional equilibrium systems where the boundaries do not affect the macroscopic phase of the system\cite{evans}.
These models have proved to be particularly useful in modeling variety of processes such as transport across biomembrane channels \cite{chou}, cellular cargo transport \cite{freylet,freypre,ignapre,madanpre,menon}, stochastic pumps\cite{jain,debasish}, fungal growth \cite{fungievans,fungiepl} as well as the collective dynamics of interacting molecular motors \cite{camps,pinkoviezky,hamid}.

 The cytoskeletal filament on which the molecular motors move is represented as a one-dimensional lattice of length $L$. The motors bound to the filament are the 'particles' in this description. The total number of lattice sites is  $N$ so that the lattice spacing $\varepsilon = L/N$. The 
occupation number of a lattice site $i$ is denoted by $n_i$ which assumes values $n_i = 1$ if occupied and  $n_i = 0$ otherwise. Particles hop unidirectionally to the right on this lattice with excluded volume interactions preventing two particles from occupying a given site as in the usual TASEP model as long the particles are free (not part of any cluster). The short range attractive interaction between the motors is incorporated by including next-nearest neighbour (NNN) interaction. Due to the attractive interaction, the motors exhibit an enhanced tendency to form a cluster while the propensity to detach from a cluster is diminished.  We include this effect by choosing a hop rate for  a free particle (not part of a cluster) to jump into a cluster as $q$, with $q>1$ while its hop rate out of motor cluster is chosen to be $r \equiv\frac{1}{q}$\cite{hamid}. Otherwise if a free particle hops to right and remains free then the hop rate is $1$. We also choose a hop rate of $1$ if a particle which is part of cluster jumps into another cluster. 
 At the filament ends, particles enter the left boundary site with a rate $\alpha$ if the particle does not join a cluster, i.e; if the neighbouring site on the right of the boundary site is unoccupied, while the entry rate is $q\alpha$  if the particle enters the boundary and the adjacent site is occupied by a particle. For the boundary site at the filament end, the particles leave the lattice with a rate $\beta$ as long the neighbouring site to the left of the right boundary site is unoccupied, while the the exit rate of particles is $\beta/q$ if the particle at boundary site is part of a cluster.
 The effect of embedding environment is incorporated by allowing for stochastic attachment and detachment process of motors on the filaments. This process breaks the particle conservation of the motors in the bulk of the filament. The introduction of Langmuir kinetic process of attachment and detachment is done  such that a particle in the bulk can detach from a lattice site with a rate $\omega_D$ while a particle could attach to a vacant lattice site with rate $\omega_A$. We restrict ourselves to the regime where the boundary processes of particle input and output compete with the bulk process of stochastic attachment and detachment. Thus we define bulk attachment rate $\Omega_{A} = \omega_{A}N$ and bulk detachment rate $\Omega_{D} = \omega_{D}N$ and hold the bulk attachment and detachment rates fixed as $N\rightarrow \infty$ \cite{freylet,freypre}.

The various processes are summarized in Fig. 1. We normalize the the total length of the lattice $L$ to $1$ .
The continuum Mean Field (MF) evolution equation of mean density of particles  is obtained by taking the expectation value of site occupation numbers, invoking the usual procedure of factorizing the two point correlators arising out of different combinations of site occupation numbers and then expressing the resultant equations in terms of continuum variable of relative position $x$ along the filament lattice \cite{freylet,freypre,ignapre}. The continuum MF evolution equation in the bulk reads as,

\begin{equation}
\frac{\partial \rho}{\partial t} = -\varepsilon \left[ \frac{\partial}{\partial x}J(\rho) +  \Omega_A(1 - \rho) - \Omega_D \rho \right]
\label{evoeqn}
\end{equation}
where the current has the form
\begin{equation}
J(\rho) = \rho(1 - \rho)[(1 - \rho)^2 + \rho^2] + (q + \frac{1}{q})\rho^2(1 - \rho)^2.
\label{current}
\end{equation}
Here we have displayed terms unto the leading order in $\varepsilon$. 

The corresponding equation for steady state density profile $\rho(x)$ is obtained by simply setting the time derivatives of $\rho$ in Eq.~\ref{evoeqn} to zero. For satisfying the left boundary condition at $x = 0$, the density profile has to satisfy the conditions $\rho(0) = \alpha$ and for density profile satisfying the right boundary condition, $\rho(1) = 1 - \beta$. 

It is worthwhile pointing out all the different limits of this model. In the absence of next nearest neighbour (NNN) interaction and Langmuir kinetics (LK) of particle attachment and detachment, the dynamics of the model reduces to the usual TASEP. In this limit, which is obtained by setting $q =1$ and $\Omega_D, \Omega_{A}=0$, we obtain steady state MF equation which corresponds to  MF steady state solution for TASEP. For the situation when LK is present, while the NNN interaction is absent, the steady state equation reduces to steady state equation for the model discussed in \cite{freylet,freypre}, wherein the steady state allows for phase coexistence and shocks localized in the bulk. Finally when the the Langmuir kinetics is absent while NNN interaction is included, then the MF steady state solution is same as  the steady state density solution discussed in \cite{popkov2}. In fact in this limit, the model maps on to Katz- Lebowitz- Spohn (KLS) model \cite{kls1,kls2}. For KLS model particle hopping is dependent on NNN and  is defined by the rules: $1 0 1 1 \rightarrow 1 1 0 1$, $1 0 1 0 \rightarrow 1 1 0 0$, $0 0 1 0 \rightarrow 0 1 0 0$ and $0 0 1 1 \rightarrow 0 1 0 1$, where $1$ corresponds to a particle occupancy at a lattice site while $0$ corresponds to a vacancy, and where these processes occur in general with different rates \cite{kls1, kls2}. In the absence of LK, the model discussed in this letter exactly maps to the dynamics of the KLS model, by  replacing particles of the KLS model by vacancies and vacancies of KLS model by particles for our case. While for KLS model it has been shown that MF analysis breaks down for strong NNN interaction, we restrict ourselves to the situation when the NNN interaction is {\it weak}, as is the case for motor interactions on cytoskeletal filaments. In this limit the MF picture can fairly accurately describe the steady state behaviour as confirmed by Monte Carlo simulations.  Further, unlike KLS model, here the interplay of the bulk translational dynamics on the lattice with the Langmuir Kinetic processes gives rise to qualitatively new scenario, wherein we find that phase coexistence and localized shocks in the bulk which are also controlled by attachment-detachment rates associated with LK process apart from the NNN coupling strength and the boundary entry and exit rates.  Thus the model discussed in this letter  can be considered as a non-conserved version of the KLS model.

 In order to simplify the analysis and construct the entire MF phase diagram in the {\it weak} motor interaction limit, we restrict our analysis for a special choice of Langmuir kinetic processes wherein the attachment rate $\Omega_{D}$ is equal to the detachment rate $\Omega_{A}$. For this special case, when $\Omega_{D} = \Omega_{A} = \Omega$, the steady state differential equation for density is,
\begin{equation}
(1-2\rho)\left[(1 - 2\rho)^2 + 2(q + \frac{1}{q})\rho(1 - \rho)\right ]\frac{d\rho}{dx} = \Omega(1-2\rho)
\label{diffrho}
\end{equation}
Like in case of TASEP, one of the possible solution of Eq.(\ref{diffrho}) is the homogeneous solution, $\rho(x) = \frac{1}{2}$. This corresponds to the maxima of current in the lattice. The corresponding expression for the current in this maximal current (MC) phase is $J_{M} = \frac{1}{8} + \frac{1}{16}(q + 1/q)$ and the density profile becomes independent of the boundary densities. Unlike TASEP, the maximal current in the lattice for this case is $J_{M} < 1/4$ for $q \neq 1$. Apart from this homogeneous solution for the density, the other possible solution is obtained by integrating Eq.~(\ref{diffrho}) and is in form of a form a cubic equation in $\rho$,
\begin{equation}
\rho^3 - \frac{3}{2}\rho^2 - \frac{3}{A}\rho + \frac{3}{A}(\Omega x + C) = 0
\label{cubic}
\end{equation}
where $A = 2(q + 1/q) - 4$ and $C$ is a constant which is fixed by appropriate boundary condition. 
Using the mean-field boundary conditions in the steady state, $\rho(0) = \alpha$ and $\rho(1) = 1 - \beta$, we can determine $C$ from Eq.~(\ref{cubic}). Substituting $C$ back in Eq.~\ref{cubic}, we can then solve for the cubic equation to determine the physically relevant solutions which will give the density profile $\rho(x)$. This is then used in Eq.~\ref{current}, to calculate the current in the different phases.
When the density profile satisfies the boundary condition at the left boundary at $x = 0$ , $\rho(0) = \alpha$ is substituted in Eq.~(\ref{cubic}) to obtain the corresponding expression for $C$ and the entire density profile in the range $0 \le x \le 1$ is subsequently obtained by solving Eq.(\ref{cubic}). This corresponds to the Low Density (LD) solution. Similarly the High Density (HD) solution for the density profile is obtained by by using the boundary condition $\rho(1) = 1 - \beta$.  It is important to note that Eq.~\ref{diffrho} is a first order differential equation which has to satisfy two boundary conditions. Therefore, the problem is overdetermined and the two solutions that respects the two boundary conditions will in general not match and give rise to a shock in the bulk. The position of the shock is determined by the continuity of the current $J$ for the two different solutions. In general three different density profiles may coexist on the filament lattice, e.g; the LD, HD and the MC phase. For any two phases to coexist, the current $J$ corresponding to the two different phases at a particular location $x = x_{o}$ on the lattice must be equal for the domain range $0 < x < 1$, $x_{o}$ being the position of the domain wall separating the two phases. While the density profile is continuous across a domain wall separating a LD or HD phase region with a MC phase region, the density change across a domain wall separating a LD phase with HD phase is discontinuous and results in a shock localized in the bulk (Fig.~\ref{3phase}). The location of the domain wall coincides with the position where the current is maximum, the current increasing and decreasing monotonically with $x$ on either side of it. With increasing interaction (increasing $q$), the domain wall shifts to the left. For very strong interaction, the domain wall disappears as the system enters into the HD phase, the density profile being larger than 1/2. As expected, the mean field results no longer hold for large $q$. 
\begin{figure*}[] 
\begin{center}
\includegraphics[width=16cm]{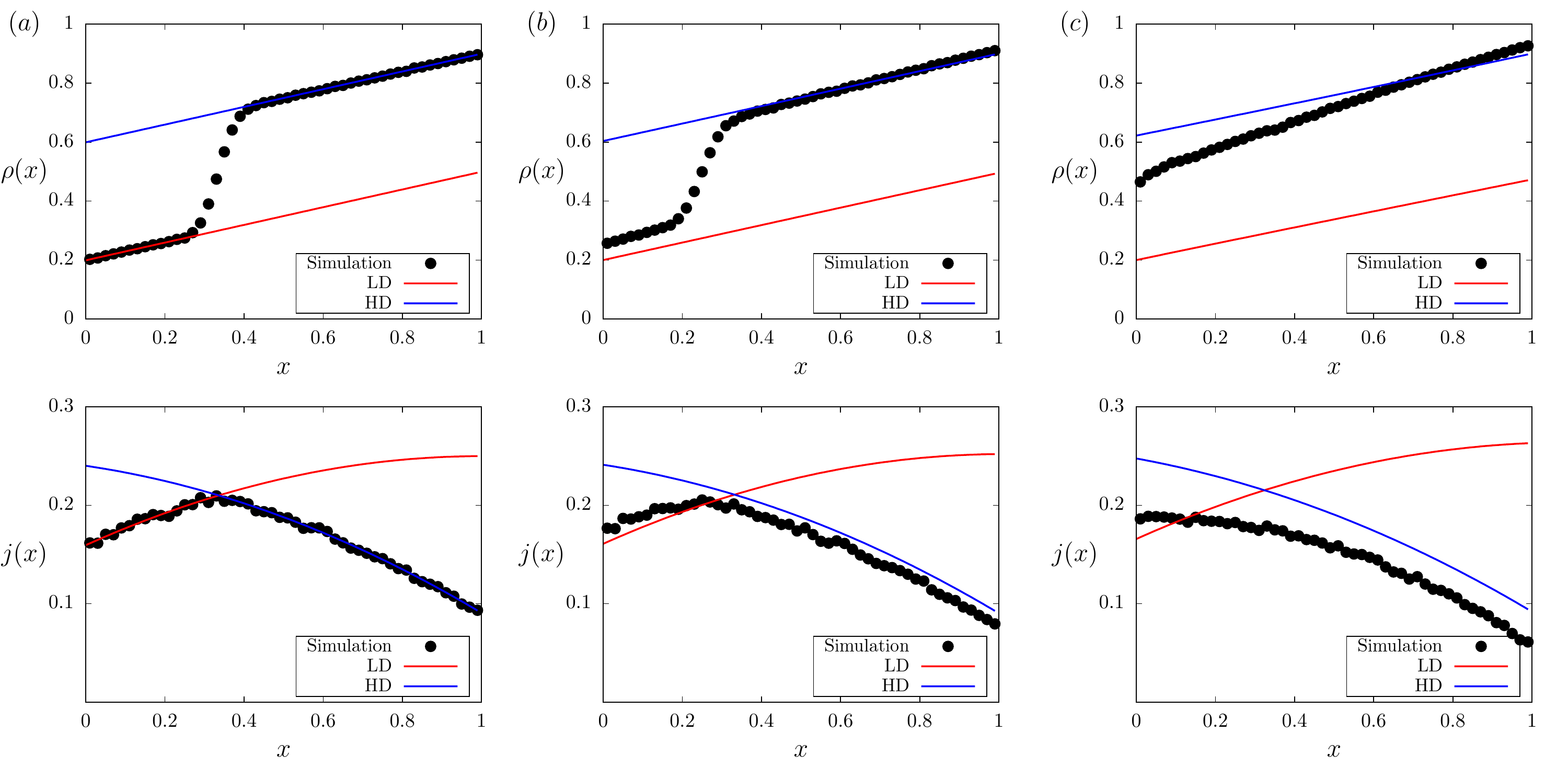} 
\caption{(color online)
Average density $\rho(x)$ and current $j(x)$ profiles for parameters $\alpha = 0.2, \beta = 0.1, \Omega = 0.3$ and (a) $q = 1.0$ (b) $q = 1.2$ and (c) $q = 1.6$. Profiles calculated analytically and from Monte Carlo simulations.}
\label{3phase}
\end{center}
\end{figure*}

\begin{figure}[h] 
\begin{center}
\includegraphics[]{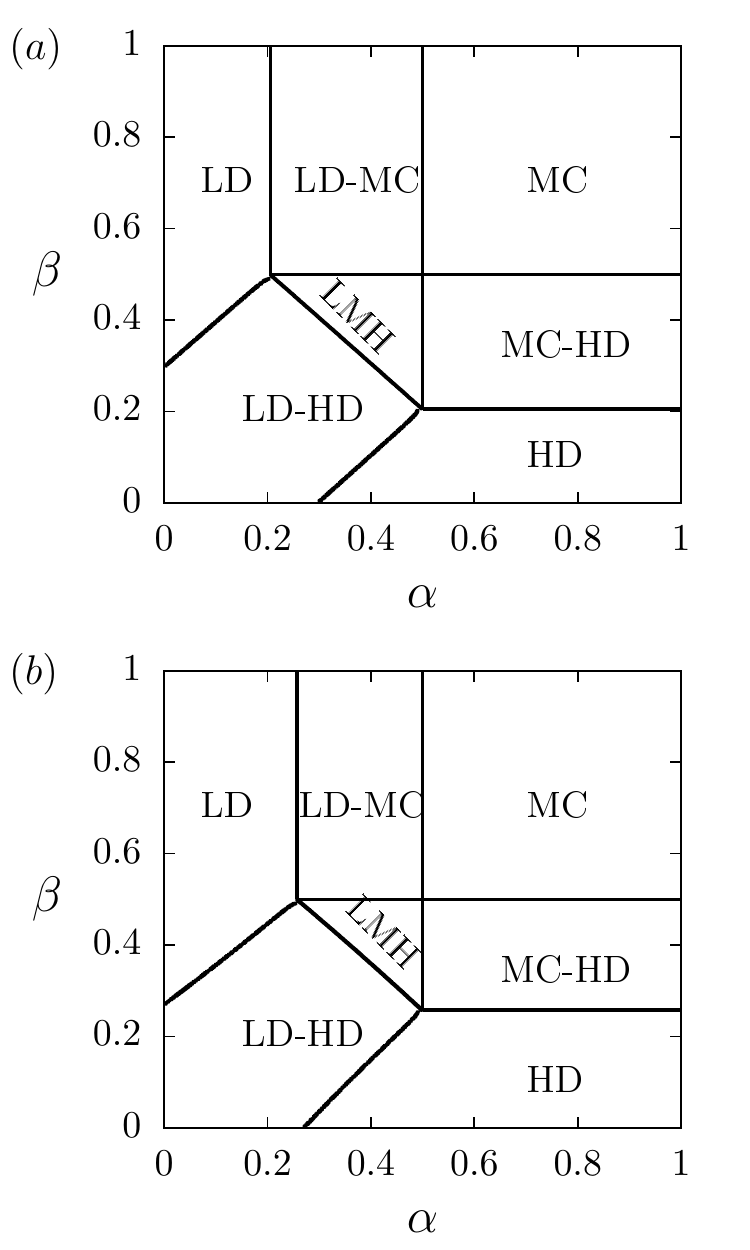} 
\caption{(color online)
Mean field phase diagram: (a) q = 1.2, $\Omega = 0.3$, (b) q = 2.0, $\Omega = 0.3$.}
\label{phasediagram}
\end{center}
\end{figure}

\begin{figure}[h] 
\begin{center}
\includegraphics[width=8cm]{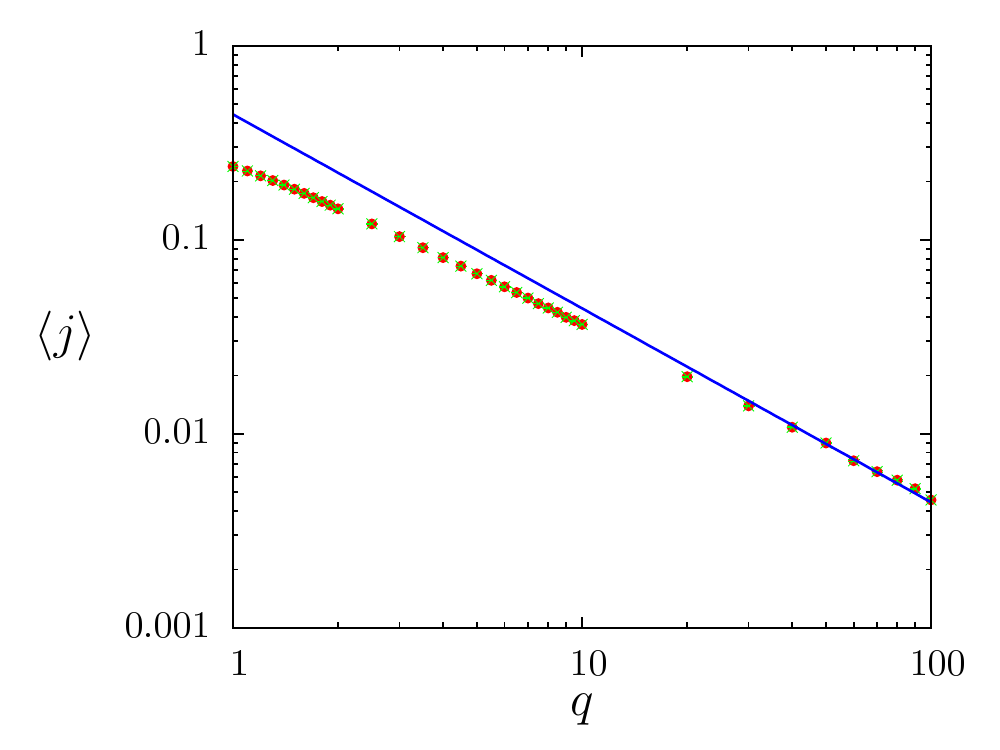} 
\caption{(color online)
Steady state current as a function of $q$ for $\alpha = 0.4, \beta = 0.4, \Omega = 0.0$. Points are simulations with error bars; line shows fit to $\simeq 1/q$ for large $q$.}
\label{phasediagram}
\end{center}
\end{figure}

In general the possible phases  in the bulk of the filament are the pure LD, HD and MC phases, two-phase coexistence of LD with MC (LD-MC), HD with MC (HD-MC) and LD with HD (LD-HD), and three-phase coexistence of LD with MC and HD ( LMH). LD-HD phase is characterized by density discontinuity and shock localized in the bulk of the lattice. The phase boundaries separating the different possible phases are determined by setting the domain wall position to left or the right end of the filament lattice.  We now discuss the conditions which determines the MF phase boundaries that separates the different possible phases.

{\it Phase coexistence boundary between LD and LD-HD}:  Starting from Eq.(\ref{cubic}), the density profile corresponding to LD phase is obtained by using the boundary condition $\rho(0) = \alpha$. The corresponding current profile $J(x)$ is obtained by using Eq. (\ref{current}). Similarly  the current solution for HD phase is obtained by using the boundary condition $\rho(1) = 1 - \beta$. To obtain the phase boundary, the current solution corresponding to LD phase $J_{LD}$  is equated with the current solution corresponding to HD phase $J_{HD}$ at $x = 1$, i.e; $J_{LD}(x = 1) = J_{HD}(x = 1)$. 

{\it Phase coexistence boundary between HD and LD-HD}:  For obtaining the phase boundary, the current solution corresponding to LD phase $J_{LD}$  is equated with the current solution corresponding to HD phase $J_{HD}$ at $x = 0$, i.e; $J_{LD}(x = 0) = J_{HD}(x = 0)$. 

{\it Phase coexistence boundary between LD and LD-MC}:  For obtaining the phase boundary in this case, the density solution for the low density phase $\rho_{LD}$ is equated to 1/2 at the right boundary, i.e; $\rho_{LD}(x = 1) = \frac{1}{2}$. The resultant phase boundary has the equation of a straight line.

{\it Phase coexistence boundary between MC and LD-MC}: Here the phase boundary is determined by equating the density solution for the LD phase $\rho_{HD}$  to 1/2 at the left boundary of the filament, i.e; $\rho_{LD}(x = 0) = \frac{1}{2}$. This is simply an equation of straight line $\alpha = 1/2$.

{\it Phase coexistence boundary between MC and HD-MC}:  For obtaining the phase boundary in this case, the density solution for the high density phase $\rho_{HD}$ is equated to 1/2 at the right boundary, i.e; $\rho_{HD}(x = 1) = \frac{1}{2}$. Thus the resultant phase boundary is a straight line $\beta = 1/2$.

{\it Phase coexistence boundary between HD and HD-MC}:  For obtaining the phase boundary in this case, the density solution for the high density phase $\rho_{HD}$ is equated to 1/2 at the left boundary, i.e; $\rho_{HD}(x = 0) = \frac{1}{2}$. The resultant phase boundary is a straight line.

{\it Phase coexistence boundary between LD-M-HD and LD-HD}: This is obtained by requiring that the density solution for the high density phase $\rho_{HD}$, low density phase $\rho_{LD}$ are simultaneously equated to  1/2 at a location $x$ in the bulk of the lattice.

In Fig. 3 we display the MF phase diagram for different values of $q$. When $q$ is sufficiently small, the topology of the phase diagram is very similar to that of TASEP-LK phase diagram as in Ref.\cite{freypre}. With increasing value of $q$ the location of the phase boundaries changes. Further unlike TASEP-LK, the phase boundary curve separating LD-HD phase with HD and the phase boundary curve separating LD-HD phase with LD is not a straight line. For high values of $q$ corresponding to strong NNN interaction, the MF predications significantly deviates from the actual density and current profiles and the corresponding phase boundaries.  In the limit of $q >> 1$, the MF picture completely breaks down. In the absence of LK process, the current in the lattice in this limit is controlled solely by the slowest process in the lattice, i.e; the rate at which the particles leave the cluster. In the absence of Langmuir kinetics, the corresponding steady state current in the lattice  $J\simeq 1/q$, for large $q$. Fig.4 shows this variation.

In summary we have studied a driven lattice gas model for the transport of molecular motors which not only interact with each other but also with cellular environment through adsorption-desorption Langmuir kinetics. We showed that the resulting phase diagram is extremely rich, with  a variety of inhomogeneous phases, including localized shocks in the bulk. Our mean field results match reasonably well with Monte Carlo simulations for weak interactions. The model maps to the non-conserved version of KLS model. For strong interactions between particles, the simple MF theory breaks down.  
\newpage
{\bf Acknowledgment}\\
SM acknowledges DBT RGYI Project No. BT/PR6715/GBD/27/463/2012 for financial support.

\end{document}